\documentclass[prb,showpacs,unsortedaddress,twocolumn,floatfix]{revtex4}

\usepackage{graphicx}
\usepackage{amsmath,amsfonts,amssymb}
\usepackage{units}
\usepackage{textcomp}

\newcommand{\V}{V$_{3}$Si}
\newcommand{\scr}{superconductor}
\newcommand{\scc}{superconducting}
\newcommand{\scon}{superconductivity}
\newcommand{\rec}{reconstructed surface area}
\newcommand{\tun}[2]{\unit[#1]{mV}, \unit[#2]{nA}}

\begin{document}

\bibliographystyle{prsty}

\title{Surface reconstruction and energy gap of superconducting V$_\mathbf{3}$Si(001)}
\author{N.\ Hauptmann}\email{hauptman@physik.uni-kiel.de}
\author{M.\ Becker}
\author{J.\ Kr\"{o}ger}
\author{R.\ Berndt}
\affiliation{Institut f\"ur Experimentelle und Angewandte Physik,
Christian-Albrechts-Universit\"at zu Kiel, D-24098 Kiel, Germany}

\date{\today}

\begin{abstract}
A yet unknown surface reconstruction of V$_3$Si(001), which is most
likely induced by carbon, is used to investigate the quasi-particle energy
gap at the atomic scale by a cryogenic scanning tunneling microscope. The
width of the gap was virtually not altered at and close to carbon impurities,
nor did it change at different sites of the reconstructed surface lattice. A
remarkable modification of the spectroscopic signature of the gap was induced, however, upon moving the tip
of the microscope into controlled contact with the superconductor. Spectroscopy
of the resulting normal-metal -- superconductor junction indicated the presence
of Andreev reflections.
\end{abstract}


\maketitle

\section{Introduction}
Superconductivity is one of the most vividly investigated research fields
in solid state physics. In particular, spectroscopy of the energy gap of the
quasi-particle density of states (DOS) has attracted considerable interest.
Modifications of the energy gap by local impurities or by passing high currents
through superconducting junctions have been recently studied.
\cite{suhl,gorkov,woolf,zittartz,ali,anderson,jth_87,ehu_01,suc_01,spa_03,scheer,cuevas,post,post_2,tinkhambook}
Effects of impurities on the energy gap
\cite{suhl,gorkov,woolf,zittartz,ali,anderson,jth_87,ehu_01,suc_01,spa_03}
provided valuable information about the interplay between superconductivity
and magnetism or about the Cooper-pairing mechanism in unconventional
superconductors, while high currents through superconducting contacts were
used to unravel the number and transmission probability of transport channels
via the occurrence of Andreev reflections. \cite{scheer, cuevas,post,post_2,tinkhambook}
Until now such experiments have been restricted to conventional elemental
superconductors with critical temperatures $T_{\text{c}}<10\,\text{K}$ or to
unconventional cuprate-based superconductors. \cite{ofi_07}

In this article we present results of a combined scanning tunneling microscopy
(STM) and spectroscopy (STS) study of the superconducting compound V$_3$Si(001),
which exhibits a yet unknown surface reconstruction. Auger electron spectroscopy
(AES) reveals that this reconstruction is most likely induced by the presence
of carbon. This surface is an ideal system to study possible local variations
of the quasi-particle energy gap since it provides, owing to the reconstruction,
carbon impurities and a spatially inhomogeneous geometry. Apart from these
favorable properties, STM investigations of this surface are extremely scarce. Except for two investigations of the magnetic flux lines\,\cite{sosolik} and Josephson tunneling\,\cite{nbe_08} on unreconstructed V$_3$Si(001) using STM and STS, to our knowledge no additional STM and STS data are available for this surface. By combining a structural analysis using STM and an
investigation of electronic properties with STS we found that the width of
the energy gap is virtually independent of the position on the surface, while
the symmetry of the gap is affected. We further investigated the current
dependence on the tip-surface distance, from which we extracted the local
apparent barrier height and the transition from tunneling to contact. At contact,
spectra of the quasi-particle energy gap exhibited pronounced modifications
compared to spectra acquired in the tunneling regime. These modifications are
discussed in terms of Andreev reflections.

We used \V{}, which belongs to the class of A$_3$B
binary intermetallic compounds or, shortly, A15 materials. \V{} exhibits the
$\beta$-tungsten structure \cite{muller} and is a type-II superconductor with
a critical temperature $T_{\text{c}}$ of \unit[17]{K}. \cite{hardy, vonsovsky}
It has interesting electronic properties \cite{clogston,morin,williams} such
as a strong temperature dependence of the magnetic and electronic susceptibility
and a large specific heat. Furthermore, \V{} undergoes
a structural cubic-to-tetragonal phase transition at a temperature of \unit[21.3]{K}, which exhibits characteristics of a thermo-elastic martensitic phase transition. \cite{paduani} The concomitant change of lattice parameters is below the detection limit of STM.\,\cite{testardi} We further notice that the occurrence of structural domain walls, which is attended by the martensitic transition is not related with the observed surface reconstruction, since the reconstruction is present already at room temperature.

The (001) surface of \V{} has attracted particular
attention because of the hexagonal-to-square transformation of its vortex lattice
structure \cite{yet1,kogan,sosolik,sonier,yethiraj} and the occurrence of surface
reconstructions. \cite{aono, zajac} The energy gap has been determined by a
variety of spatially averaging techniques
\cite{nefyodov,levinstein,hauser,schumaun,moore,bangert,tanner,hackl,reinert}
yielding two distinct values of the zero-temperature energy gap $\Delta_0$
either in the range from $1.32\,\text{k}_{\text{B}}\,T_{\text{c}}$ to
$2.7\,\text{k}_{\text{B}}\,T_{\text{c}}$ or in the range from
$0.5\,\text{k}_{\text{B}}\,T_{\text{c}}$ to
$0.95\,\text{k}_{\text{B}}\,T_{\text{c}}$. The existence of these two distinct
values of $\Delta_0$ has so far been explained by BCS theory in terms of
overlapping bands.\cite{nefyodov,suhlbcs,moskalenko} The first range of energy
gaps is in good agreement with the value predicted by single-band {\em s}-wave
BCS theory ($\Delta_0=1.76\,\text{k}_{\text{B}}\,T_{\text{c}}$). \cite{bcs}

\section{Experiment}
Measurements were performed with scanning tunneling microscopes operated in
ultrahigh vacuum with a base pressure of $10^{-9}\,\text{Pa}$ and optimized
for low temperatures of $(7.5\pm 0.3)\,\text{K}$ and for room temperature.
Chemically etched tungsten tips were further prepared {\it in vacuo} by annealing
and argon ion bombardment, while gold tips were cut at ambient conditions
and used for the experiment without further treatment. The V$_3$Si(001) surface
of a polished single crystal was prepared by argon ion bombardment with ion
energies between $0.5$ and $2\,\text{keV}$ and subsequent annealing at
temperatures between $1100$ and $1200\,\text{K}$. Temperature readings from
the pyrometer used in the experiments have an uncertainty margin of
$\approx\!\pm\,50\,\text{K}$. All STM images were acquired in the constant-current
mode with the voltage applied to the sample. STS measurements were performed
in the constant-height mode. Spectra of the differential conductance
($\text{d}I/\text{d}V$) were acquired by superimposing a sinusoidal voltage
signal (root-mean-square amplitude $1\,\text{mV}$, frequency $7.6\,\text{kHz}$)
onto the tunneling voltage and by measuring the current response with a
lock-in amplifier. For AES a cylindrical mirror analyser was used.

\section{Results and discussion}
\subsection{Characterization of reconstructed \V{}(001)}
\label{modell_reconstruction}
Depending on the preparation parameters, $(2\times 1)$ reconstructed
or $(1\times 1)$ unreconstructed surfaces of V$_3$Si(001) have been reported
previously by  Zajac {\it et al.} \cite{zajac} They found by low-energy
electron diffraction and AES that annealing of V$_3$Si(001) at temperatures
exceeding $1070\,\text{K}$ leads to a $(2\times 1)$ reconstruction, while the
$(1\times 1)$ substrate surface is obtained for annealing temperatures lower
than $1020\,\text{K}$.
\begin{figure}
  \includegraphics[width=85mm,clip=]{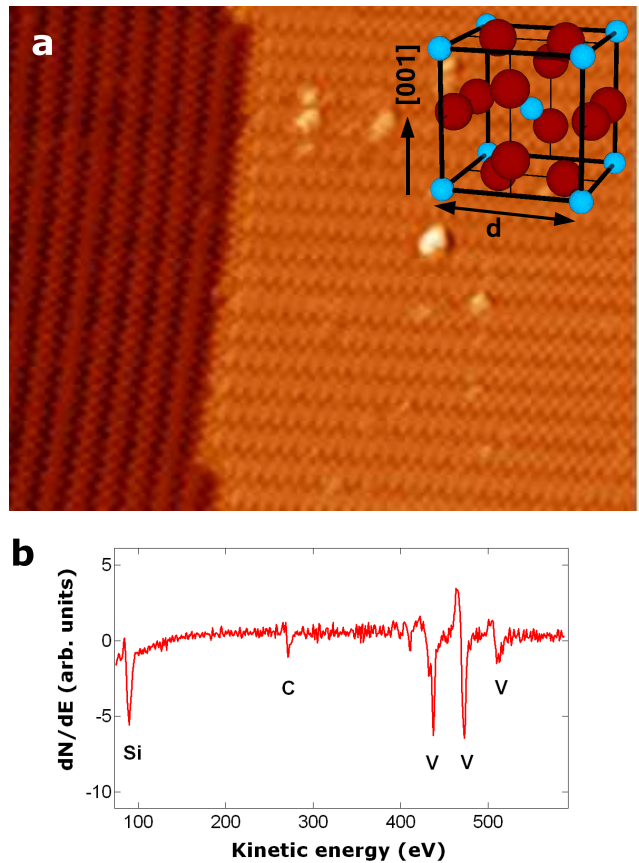}
  \caption[Fig:large_area]{(Color online) (a) STM image of two adjacent terraces
  on V$_3$Si(001) at room temperature
  (\tun{420}{2.4}, \unit[50]{nm}\,$\times$\,\unit[40]{nm}). The parallel lines
  visible on the terraces are due to a reconstruction of the surface, which
  is discussed in the text. Inset: $\beta$-tungsten lattice structure
  (Small filled spheres: silicon atoms, large filled spheres: vanadium atoms). The arrow indicates the [001] direction. (b) Auger spectrum of the reconstructed \V{}(001) surface acquired with a kinetic energy of impinging electrons of \unit[3]{kV}. Silicon (\unit[92]{eV}), carbon (\unit[274]{eV}), and vanadium (\unit[439]{eV}, \unit[475]{eV}, \unit[512]{eV}) peaks are identified according to Ref.\,\onlinecite{auger}.}
  \label{Fig:large_area}
\end{figure}
Figure \ref{Fig:large_area}(a) shows a room-temperature STM image of a V$_3$Si(001) surface
obtained using the preparation procedure introduced in the experimental
section. The reconstruction observed here does not match the previously
observed (2\,$\times$\,1) or (1\,$\times$\,1) meshes.

Terraces which exhibit parallel lines are observed. The corrugation
of these lines is $(0.06\pm 0.02)\,\text{nm}$ and almost independent
of the voltages applied in the experiments ($|V|\leq
1.5\,\text{V}$). The distance between two adjacent lines is
\unit[(1.8\,$\pm$\,0.1)]{nm}, which is approximately four times the
lattice constant of \V{}, $d=$\,\unit[0.47]{nm}. The lines on
adjacent terraces are either parallel [not shown in
Fig.\,\ref{Fig:large_area}(a)] or orthogonal with respect to each
other. The step height between adjacent terraces with orthogonal
orientation of the lines is $(0.23\pm 0.05)\,\text{nm}$, which
corresponds to half a lattice constant of V$_3$Si. Step heights
between adjacent terraces with a parallel orientation of the lines
are $(0.47\pm 0.05)\,\text{nm}$, which is similar to the V$_3$Si
lattice constant. Based on these observations we can tentatively
identify the origin of the lines. To this end we refer to the inset
of Fig.\,\ref{Fig:large_area}(a), which depicts a model of the \V{}
lattice structure. Small filled spheres represent silicon atoms
whereas large filled spheres represent vanadium atoms. The lattice
constant $d$ is identical with the spacing of the silicon
sublattice. On an unreconstructed \V{}(001) surface, vanadium atoms
would form lines on the surface of a single crystal. According to
this structural model, adjacent terraces with a step height of half
a lattice constant exhibit lines of vanadium atoms, which are
oriented perpendicular to each other. Analogously, terraces
separated by one lattice constant show vanadium atom lines with the
same orientation. This behavior matches our experimental observation
and suggests that the lines originate from vanadium atoms. The
mutual distance of the vanadium lines of $\approx 4d$, however, does
not match the separation expected for the clean crystal.
\begin{figure}
  \includegraphics[width=85mm,clip=]{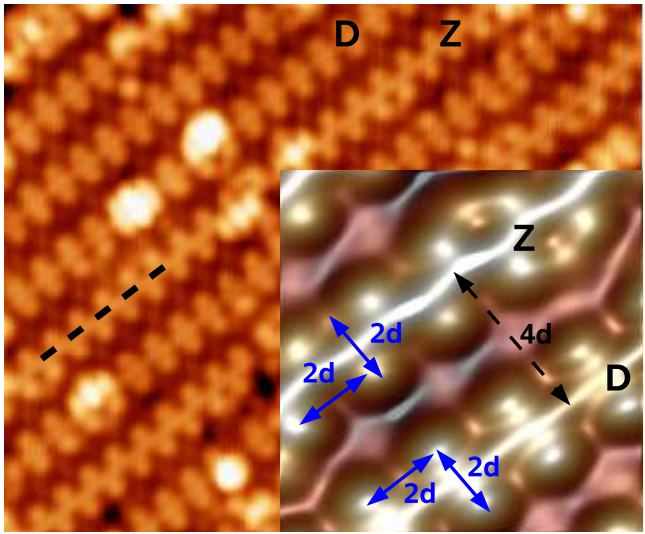}
  \caption[Fig:middle_view]{(Color online) Atomically resolved STM image of
  a V$_3$Si(001) terrace at \unit[7.3]{K} (\tun{420}{1},
  \unit[18]{nm}\,$\times$\,\unit[15]{nm}). Two characteristic lines, a dimer
  (D) and a zigzag (Z) line, appear on the terraces. Inset: Pseudo-three-dimensional presentation of close-up view
  (\unit[4]{nm}\,$\times$\,\unit[4]{nm}) showing adjacent D and Z lines. The
  distance indicated by the solid arrows is
  \unit[(0.9$\,\pm$\,0.1)]{nm}\,$\approx 2d$ where $d$ denotes the lattice
  constant, while the lines are separated by
  \unit[(1.8\,$\pm$\,0.1)]{nm}\,$\approx 4d$ (dashed arrow).}
  \label{Fig:middle_view}
\end{figure}

To obtain more information about the reconstructed surface we used
AES, which enables access to the chemical composition of the surface
within the 5 top-most atomic layers depending on the
material.\,\cite{chang} An Auger spectrum of the reconstructed \V{}
surface is shown in Fig.\,\ref{Fig:large_area}(b). In addition to
spectroscopic signatures originating from vanadium (\unit[439]{eV},
\unit[475]{eV}, \unit[512]{eV}) and silicon (\unit[92]{eV}), a
contribution from carbon (\unit[274]{eV}) is observed. Following the
quantitative analysis of Auger spectra exposed in
Ref.\,\onlinecite{auger}, which takes into account the
energy-dependent transmission function of the electron analyzer and
the element-specific Auger sensitivity factors, the probed sample
volume contains $\approx$\,\unit[8\,]{\%} with respect to the total
amount of detected elements. Taking into account the inelastic mean
free path of carbon Auger electrons, we estimated that substrate
layers within three to four lattice constants contribute
significantly to the signal, so that the amount of carbon of
$\approx$\,\unit[8\,]{\%} gives the relative carbon concentration
close to the surface. After excluding several other carbon sources
such as the argon gas used for ion bombardment, the sample holder
and the experimental setup, we surmise that carbon segregates from
the bulk to the surface. Similar carbon-induced reconstruction like
the one reported here have been reported for Si(001) and
W(110).\,\cite{derycke,yu,tejeda,bode} In the latter case, carbon
atoms arrange in a zigzag pattern, which is similar to our
observations.

Figure\,\ref{Fig:middle_view} shows an STM image of a \rec{} at
\unit[7.3]{K} in more detail. Two types of lines can be observed.
The first type of lines shows circular protrusions which are
arranged in a zigzag (Z) pattern on both sides of the line (dashed
line). The same protrusions occur on opposite sites of the second
type of lines and form a dimer (D) arrangement. We ascribe these
protrusions to carbon atoms, which are arranged on opposite sites of
the vanadium lines. The dimer and zigzag lines do not appear in an
alternating pattern. We observed, however, that a carbon atom of a
dimer line always faces a carbon atom of an adjacent line along a
perpendicular direction with respect to the vanadium lines. This
behavior may indicate an attractive interaction between the atoms.
Some impurities are observed, which appear as bright protrusions.
The inset of Fig.\,\ref{Fig:middle_view} shows a close-up view of a
dimer and a zigzag line. The distance between two carbon atoms in a
dimer line (solid arrows in the inset of
Fig.\,\ref{Fig:middle_view}) is \unit[(0.9$\,\pm$\,0.1)]{nm}, which
is approximately two times the lattice constant of \V{}. The same
distance is obtained for zigzag lines if the carbon atoms on one
side of a zigzag line are shifted by $d$ along this line. Thus, a
zigzag line can be obtained by shifting the carbon atoms on one side
of a dimer line by the lattice constant. The fact that two carbon
atoms have a distance of roughly twice the lattice constant
indicates that carbon atoms occupy every second lattice site of the
silicon sublattice. Taking into account the martensitic transition
of \V{}, the question may arise whether structural domain walls may
be the origin of the reconstruction. However, the same
reconstruction was observed at room temperature
[Fig.\,\ref{Fig:large_area}(a)] and we conclude that the martensitic
transition is not related with the observed surface reconstruction.
\begin{figure}
  \includegraphics[width=88mm,clip=]{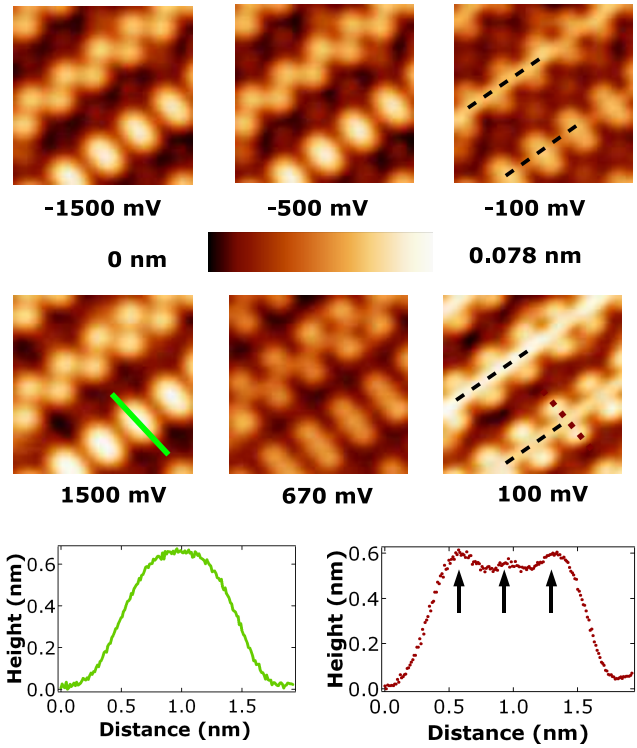}
  \caption[Fig:voltage]{(Color online) STM images of a \rec{}
  (\unit[1]{nA}, \unit[4]{nm}\,$\times$\,\unit[4]{nm}) taken at indicated
  voltages between \unit[-1500]{mV} and \unit[1500]{mV} at \unit[7.3]{K}. Two
  cross-sectional profiles of a dimer line are shown (positions are indicated
  by a solid and dotted line, respectively).}
  \label{Fig:voltage}
\end{figure}

STM images of the same reconstructed surface area at several different tunneling
voltages are shown in Fig.\,\ref{Fig:voltage}. Tip changes in between the
acquisition of the different images can be excluded since all images
were taken successively within \unit[3]{min}. Furthermore, sudden changes in
the tunneling current, which may indicate tip changes, were not observed.
The vanadium lines become pronounced at small voltages ($\pm$\,\unit[100]{mV}),
which is indicated by dashed black lines, whereas they are hardly observed at
high positive and negative voltages. This may indicate a one-dimensional
state at low energies around the Fermi energy, which is located on the lines.
Two findings lend support to this idea. First, a one-dimensional state
is in agreement with an electronic model for the DOS of \V{}.
\cite{clogston1,wegner,labbe} This model assigns quasi-one-dimensional character
to the vanadium $d$ electrons, which gives rise to sharp peaks in the DOS close
to the Fermi energy. First-principles calculations of the bulk band structure
\cite{mattheiss, mattheiss_2, klein, pui, mattheiss_3, bisi} also exhibit an
extremely narrow peak in the DOS right below the Fermi energy, which is mainly
due to vanadium $d$ electrons. We note, however, that the calculations
consider the bulk electronic structure, which may be modified at the surface.
As a second indication favoring the presence of a one-dimensional state we
mention the observation and calculation of a quasi-one-dimensional state on
Fe(100) with segregated carbon. \cite{gpa_06,gtr_06} Similar to our observations
carbon atoms form zigzag chains on Fe(100) and laterally confine Fe $s-d$
states, which gives rise to formation of a Fe surface state band with
one-dimensional character near the Fermi level.

At small positive voltages (\unit[100]{mV}) the dimers in the dimer lines can
clearly be distinguished as two atoms whereas at high negative and positive
voltages ($\pm$\,\unit[1500]{mV}) only a smeared shape can be seen. Two line
scans on a dimer line are shown in Fig.\,\ref{Fig:voltage}, which were taken
at positions indicated by the solid and dotted line, respectively. At \unit[100]{mV}
(right line scan in Fig.\,\ref{Fig:voltage}), the line and the two carbon atoms
can be clearly distinguished (black arrows), while at \unit[1500]{mV} (left
line scan in Fig.\,\ref{Fig:voltage}), only a smeared shape can be seen. This
loss of resolution can be understood on general grounds. Assuming a
constant electronic structure of the STM tip, the tunneling current is an integral over
the sample local density of states multiplied by a barrier transmission
factor. The higher the tunneling voltage, the larger is the number of states
contributing to the tunneling current. As a consequence, the localized one-dimensional state
is dominated by other states. Only imaging at low voltages (here
at \unit[$\pm$\,100]{mV}) gives rise to the sharp and localized lines along
the dashed lines in Fig.\,\ref{Fig:voltage}. Further, some protrusions between
the lines are observed (with a mutual distance of $2d$), which become more
pronounced at negative voltages. These protrusions are assigned to silicon atoms.
\begin{figure}
  \includegraphics[width=88mm,clip=]{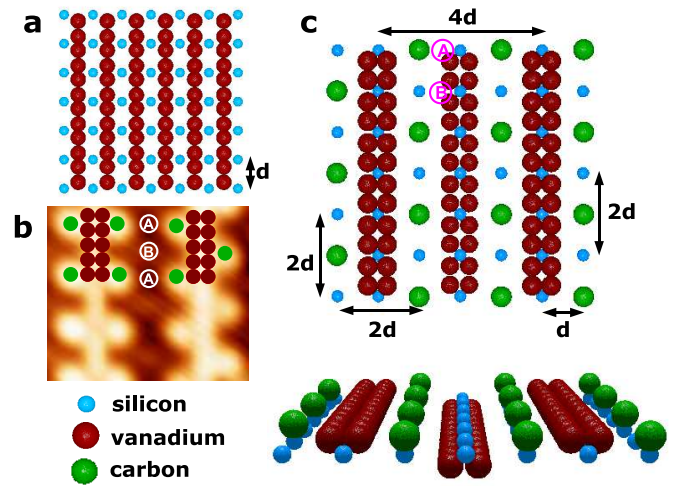}
  \caption[Fig:surface_model]{(Color online) (a) Top-most surface layer of
  unreconstructed \V{}(001) surface (small filled spheres: silicon atoms, large filled spheres: vanadium atoms). (b) STM image
  (\unit[100]{mV}, \unit[1]{nA}) illustrating the model. Filled spheres denote carbon and vanadium atoms. (c) Top and side view of the top-most surface layer of the model of the \rec{}s showing
  a dimer line and a zigzag line. A and B denote silicon atoms embedded in
  different environments regarding the neighbor atoms.}
  \label{Fig:surface_model}
\end{figure}

According to these experimental findings concerning the chemical composition ($\approx$\,\unit[8\,]{\%} carbon), the topographic (mutual atom distances),
and electronic properties (chains of vanadium atoms) of the investigated surface as outlined above, a structural model for the \rec{}s
is proposed in the following. The top-most unreconstructed \V{}(001) surface
layer is shown in Fig.\,\ref{Fig:surface_model}(a), where silicon and vanadium atoms are
indicated by small and large spheres, respectively. Two neighboring vanadium lines of
the top-most unreconstructed surface layer are proposed to
form double rows [Fig.\,\ref{Fig:surface_model}(c)]. Every other
double row of vanadium moves below the silicon sublattice into the
second crystal layer whereas the other vanadium double rows move slightly above
the silicon sublattice.
The distance between the top double rows is $4d$, matching the distance
extracted from STM images. Carbon atoms are attached to the lines on top
of some silicon atoms. Dimer lines as well as zigzag lines are formed in this
way. The distance between two carbon atoms in a dimer line in this model is $2d$.
This distance and also the corresponding distance in the zigzag lines is in
agreement with the measurements (compare with the inset in
Fig.\,\ref{Fig:middle_view}). Figure\,\ref{Fig:surface_model}(b) shows an STM image of a \rec{} where carbon and vanadium atoms are depicted by spheres. Silicon atoms denoted A and B in
Fig.\,\ref{Fig:surface_model}(c) are embedded
in different environments regarding the neighbor atoms. They are either
neighbored by two carbon atoms and four underlying silicon atoms (position A)
or by just four silicon atoms (position B). Only the latter silicon atoms
can be observed in the \rec{}s as protrusions between the lines [position B
in Fig.\,\ref{Fig:surface_model}(b)].

\subsection{Superconducting energy gap}
\label{superc}
\begin{figure}
  \includegraphics[width=85mm,clip=]{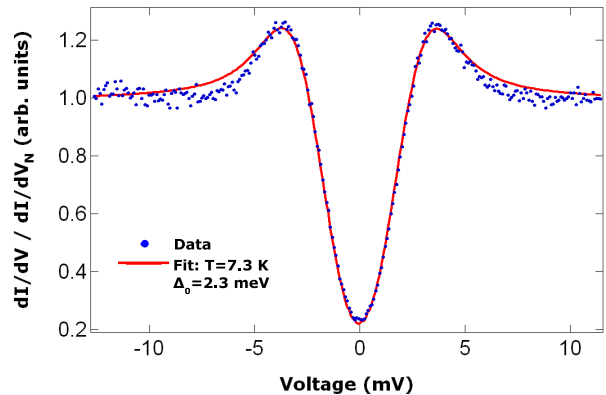}
  \caption[Fig:sc_fit]{(Color online) Fit of the BCS quasiparticle DOS (solid
  line), broadened by the temperature and the modulation voltage used for
  lock-in amplifier detection of the current response, to normalized
  $\text{d}I/\text{d}V$ spectrum (dots). The feedback loop was opened at \unit[12]{mV} and \unit[0.5]{nA}.}
  \label{Fig:sc_fit}
\end{figure}

\V{} is a conventional intermediate-coupling \scr{}. \cite{vonsovsky, poole}
The DOS of the quasi-particles for intermediate-coupling is described according
to Eliashberg.\cite{eliashberg} For simplicity, in this work we assume the energy gap to be
solely temperature-dependent, $\Delta=\Delta(T)$, such that the quasi-particle DOS
can be written as the lifetime-broadened BCS expression for the quasi-particle DOS,
$\rho_{\text{BCS}}$, given by \cite{dynes,alexandrov}
\begin{equation}
  \rho_{\text{BCS}}(E,T)= \rho_{\text{N}}(E)\,\text{Re}\left\{\frac{|E|-i\Gamma}{\sqrt{(E-i\Gamma)^2-|\Delta(T)|^2}}\right\},
  \label{DOS_intermediate}
\end{equation}
where $\rho_{\text{N}}(E)$ is the DOS of normal conducting electrons and
$\Gamma$ is a lifetime parameter. We consider a
broadening due to the temperature as well as a broadening due to the lock-in
amplifier by convoluting $\rho_{\text{BCS}}$ with the temperature-broadening
function as well as with the instrumental broadening function of the lock-in
amplifier.\cite{klein_inst,kroeger} We assume the lifetime-broadening to be
of the order of \unit[0.1]{meV}\,\cite{dynes} whereas the temperature-broadening
is around \unit[2.2]{meV} full width at half maximum (FWHM) at \unit[7.3]{K}.
\cite{kroeger} Therefore, for our fits the lifetime parameter is set to a small
value such that increasing $\Gamma$ by a factor $2$ does not change our results.
Thus only the temperature $T$ and the energy gap $\Delta$ are used as fit
parameters. The $\text{d}I/\text{d}V$ spectra close to the Fermi energy are
normalized to the differential conductance at \unit[$-$12]{mV}
($\text{d}I/\text{d}V$$_{\text{N}}$), which is well outside the energy gap.
The current at \unit[12]{mV} was set to values between \unit[0.5]{nA} and
\unit[1]{nA}. A typical fit to the data (dots) is shown as a solid line in
Fig.\,\ref{Fig:sc_fit}. Best fits are obtained for temperatures between
\unit[7.2]{K} and \unit[7.8]{K}, which are in good agreement with the temperature
of the experiment. Our fits yield energy gaps $\Delta=(2.1\,\pm\,0.2)\,\text{meV}$
within this temperature range. According to BCS theory, this corresponds to
a maximal energy gap at \unit[0]{K} of
\unit[$\Delta_0=(2.15\,\pm$\,0.20)]{meV} $=(1.47\,\pm\,0.14)\,\text{k}_{\text{B}}\,T_{\text{c}}$
with $T_{\text{c}}=$\,\unit[17]{K}. These results are in reasonable accordance
with previous measurements.
\cite{nefyodov,levinstein,hauser,schumaun,moore,bangert,tanner,reinert} However, comparing our results with most recent measurements\,\cite{nefyodov,reinert} we find slightly lower gap widths, which may be related to structural changes at the surface due to the carbon-induced reconstruction.

\begin{figure}
  \includegraphics[width=85mm,clip=]{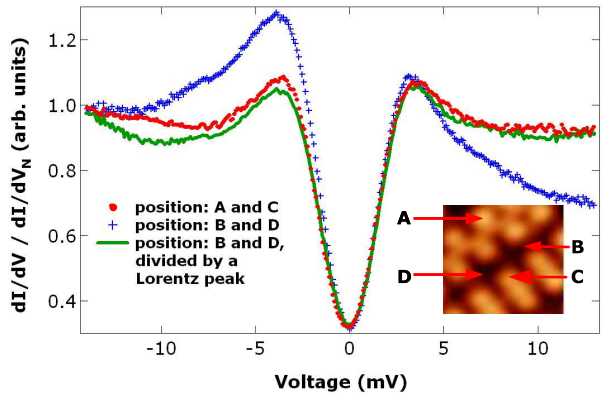}
  \caption[Fig:local]{(Color online) Spectra of $\text{d}I/\text{d}V$ at
  constant height close to the Fermi energy at different positions on the
  \rec{}s ($\bullet$: position A and C, +: position B and D, positions are denoted in the inset). The feedback loop was opened at \unit[14]{mV} and \unit[0.9]{nA}. Solid line: Spectrum
  at positions B and D after subtraction of a Lorentzian with
  a maximum position at \unit[$-5$]{meV} and FWHM of \unit[29]{meV}.}
  \label{Fig:local}
\end{figure}

To probe a local effect on $T_{\text{c}}$ we acquired
$\text{d}I/\text{d}V$ spectra at different positions of the reconstructed surface,
which are defined in the inset of Fig.\,\ref{Fig:local}.
Normalized spectra of $\text{d}I/\text{d}V$ were taken either on atoms of the
dimer or zigzag pattern (positions A and C) or between the lines (positions
B and D).
Differences between $\text{d}I/\text{d}V$ spectra taken at positions A and C
and taken at positions B and D are hardly detectable, so that mainly
two different $\text{d}I/\text{d}V$ curves can be observed, which are shown
in Fig.\,\ref{Fig:local}. Spectra of $\text{d}I/\text{d}V$ taken at positions
A and C (dots) are symmetric with respect to zero voltage whereas
those at positions B and D (crosses) are asymmetric. For small negative sample
voltages, the differential conductance is higher at positions
B and D than at positions A and C. This behavior may be due to the DOS of normal
conducting electrons $\rho_{\text{N}}$, which is different for positions A
and C (on the lines) and positions B and D (between the lines).
Typically, $\rho_{\text{N}}$ is treated as constant, which leads to a
symmetric $\rho_{\text{BCS}}$ around the Fermi energy. Since the atomic
environment of atoms at different positions of the \rec{}s is different in
the present case, it is likely that $\rho_{\text{N}}$ varies locally, too.
If $\rho_{\text{N}}$ has asymmetric features within a small energy range around
the Fermi energy, the energy gap can be asymmetric. As it can be seen from
Eq.\,(\ref{DOS_intermediate}), the influence of $\rho_{\text{N}}$ can be
cancelled by dividing by $\rho_{\text{N}}$. An angle-resolved photoemission
study of V$_3$Si(001) performed at room temperature revealed the existence
of a remarkably sharp peak close to the Fermi level and in the center of the
surface Brillouin zone. \cite{aono} This contribution to the density of states
of normal conducting electrons may give rise to the observed asymmetry of the
measured energy gap in spectra of $\text{d}I/\text{d}V$.
In a first approximation, we divided the asymmetric spectrum by a Lorentzian (maximum
at \unit[$-$5]{meV} and FWHM of \unit[29]{meV}) in order to remove the asymmetry.
The result is depicted in Fig.\,\ref{Fig:local} by the solid curve. The asymmetric
behavior is nearly compensated and the solid curve fits well with the symmetric $\text{d}I/\text{d}V$ spectrum.

\subsection{Contact spectroscopy}
To perform contact spectroscopy the tip had to be moved controllably into
contact with the V$_3$Si(001) surface. For this purpose we proceeded as
reported in Ref.\,\onlinecite{lli_05} and monitored the conductance ($G=I/V$,
$I$: current, $V$: voltage) over a range of tip displacements comprising tunneling
and contact regimes. Unlike on noble metal surfaces \cite{lli_05,nne07a,jkr_07}
the contact to V$_3$Si(001) is not always signaled by a jump of the conductance.
We also observed a continuous transition from the tunneling to the contact
regime. Contact conductances are defined according to the procedure exposed
in Ref.\,\onlinecite{jkr_08}. Contact conductances defined in this way scatter
between $\approx 1\,\text{G}_0$ and $\approx 3\,\text{G}_0$ where
$\text{G}_0=2\text{e}^2/\text{h}$ denotes the quantum of conductance
(h: Planck's constant). Once the contact regime is reached the tip could be
moved further into the surface increasing the conductance to up to
$\approx 6\,\text{G}_0$ without destroying the sample surface and tip integrity. From conductance curves acquired in the tunneling regime we inferred the
apparent barrier height within a one-dimensional model of the tunneling junction.
\cite{simmons,lang,olesen} The apparent barrier height ($\Phi_{\text{ap}}$) was determined by using a tungsten tip and also by using a gold tip. From our contact measurements we obtained
$\Phi^{\text{W}}_{\text{ap}} = $\unit[(4.60$\,\pm\,$0.5)]{eV} and
$\Phi^{\text{Au}}_{\text{ap}} = $\unit[(4.19$\,\pm\,$0.2)]{eV}. In the simplest case, the work function of the sample (denoted $\Phi^{\text{V$_3$Si}}$) is given by
$\Phi^{\text{V$_3$Si}}=2\,\Phi_{\text{ap}}-\Phi^{\text{W/Au}}$. Taking into
account the work functions for polycrystalline specimen of W,
$\Phi^{\text{W}}=\unit[4.55]{eV}$, and Au, $\Phi^{\text{Au}}=\unit[5.1]{eV}$, from Ref.\,\onlinecite{michaelson}, the work function of \V{} is
$\Phi^{\text{V$_3$Si}}=\unit[(3.96\pm 0.90)]{eV}$. Our result is in agreement
with previous measurements of the work function of \V{}, which yield
\unit[4.4]{eV} for single-crystalline \V{}(001)\,\cite{aono} and \unit[4.6]{eV}
for a polycrystalline sample. \cite{heiniger}
\begin{figure}
  \includegraphics[width=85mm,clip=]{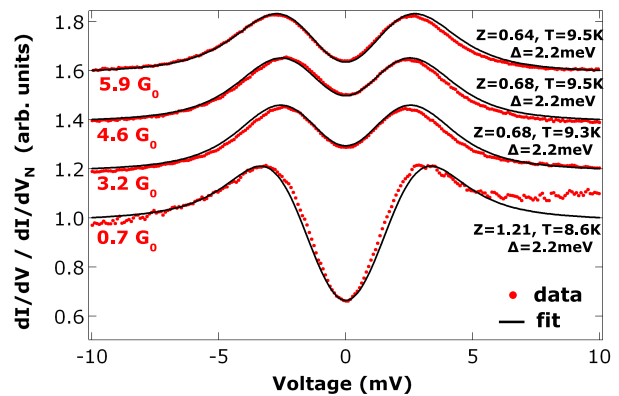}
  \caption[Fig:contact]{Spectra of differential conductance
  ($\text{d}I/\text{d}V$) for different junction conductances. The spectra
  are normalized to the conductance value at \unit[-14]{mV}
  ($\text{d}I/\text{d}V$$_N$) and are shifted vertically by
  \unit[0.2]{arb.\,units} with respect to each other. The feedback loop was opened at \unit[14]{mV} and \unit[0.7]{$\mu$A}, \unit[3.5]{$\mu$A}, \unit[5.0]{$\mu$A}, and \unit[6.5]{$\mu$A}. Fits (solid lines), according to Ref.\,\onlinecite{blonder}, and fit parameters are shown.}
  \label{Fig:contact}
\end{figure}

For contact spectroscopy we proceeded as follows. We first acquired a
conductance-versus-displacement curve in order to find the contact regime.
We then set a junction conductance corresponding to tip-surface contact and
performed spectroscopy of $\text{d}I/\text{d}V$ as previously reported in
Refs.\,\onlinecite{nne07a,nne_08,nne_09}. Typical results for various junction
conductances are shown in Fig.\,\ref{Fig:contact}. The $\text{d}I/\text{d}V$
spectra are normalized to the conductance value at \unit[-14]{mV} and are
shifted vertically with respect to each other. The energy gap in the
$\text{d}I/\text{d}V$ spectra becomes less pronounced at higher conductances
of the junction, {\it i.\,e.}, the indentation of the $\text{d}I/\text{d}V$
curves between the condensation peaks becomes shallower upon increasing the
junction conductance. To model this behavior we first tried to fit the BCS
density of states with an increased effective temperature to the data. Several
contact experiments have been reported where local heating of the contact
junction due to a high current density plays a significant role.
\cite{nne07b,gsc_08} However, increasing the effective temperature in the BCS
density of states, Eq.\,(\ref{DOS_intermediate}), decreases the condensation
peaks, which is not observed in our data.

The apparent change of the \scc{} energy gap with increasing junction conductance may be
explained by taking Andreev reflections into account.
\cite{benistant, soulen, strijkers, andreev} To fit our data,
we use a theoretical approach to Andreev reflections at an interface between
a normal-metal and a \scr{} reported by Blonder {\it et al.}\,\cite{blonder}
These authors introduced a $\delta$-function potential barrier of strength
$Z$ to describe the normal-metal -- superconductor interface. For a large
barrier strength ($Z=3.0$) calculated $\text{d}I/\text{d}V$ spectra agree
with $\text{d}I/\text{d}V$ data acquired in the tunneling regime
(Fig.\,\ref{Fig:sc_fit}). We used the energy gap, $\Delta$, the temperature,
$T$, and the dimensionless barrier strength, $Z$, as fitting parameters. Fits
for the spectra at different conductances are shown in Fig.\,\ref{Fig:contact}
by solid lines. The quality of the fits is quite remarkable indicating that Andreev
reflections play a role in our contact spectroscopy of the V$_3$Si(001)
energy gap.

\section{Summary}
STM of V$_3$Si(001) reveals a surface reconstruction, which is most likely
induced by carbon. Based on atomically resolved STM images and on the chemical
composition of the surface as monitored by AES we suggest a structural model
of this reconstruction. Spatially resolved spectroscopy revealed that the width
of the energy gap does not depend on the surface site, while the symmetry of the gap is affected
depending on the local environment. The spectroscopic signature of the energy gap changes in the contact regime. These modifications are compatible with Andreev reflections in the point contact between a normal metal tip and a superconductor.

We thank M.\ Ternes for discussions and the Deutsche Forschungsgemeinschaft
for financial support through KR 2912 3-1.


\begin{thebibliography}{nnnyys}

%INTRODUCTION
\bibitem{suhl} B.\ T.\ Matthias, H.\ Suhl, and E.\ Corenzwit,
\prl {\bf 1}, 92 (1958).

\bibitem{gorkov} A.\ A.\ Abrikosov and L.\ P.\ Gorkov,
Sov.\ Phys.\ JETP {\bf 12}, 1243 (1960).

\bibitem{woolf} M.\ A.\ Woolf and F.\ Reif,
Phys.\ Rev.\ {\bf 137}, A557 (1965).

\bibitem{zittartz} J.\ Zittartz, A.\ Bringer, and E.\ M\"uller-Hartmann,
Solid State Commun.\ {\bf 10}, 513 (1972).

\bibitem{ali} A.\ Yazdani, B.\ A.\ Jones, C.\ P.\ Lutz, M.\ F.\ Crommie, and
D.\ M.\ Eigler,
Science {\bf 275}, 1767 (1997).

\bibitem{anderson} P.\ W.\ Anderson,
J.\ Phys.\ Chem.\ Solids {\bf 11}, 26 (1959).

\bibitem{jth_87} J.\ R.\ Thompson, D.\ K.\ Christen, S.\ T.\ Sekula,
B.\ C.\ Sales, and L.\ A.\ Boatner,
\prb {\bf 36}, 836 (1987).

\bibitem{ehu_01} E.\ W.\ Hudson, K.\ M.\ Lang, V.\ Madhavan, S.\ H.\ Pan,
H.\ Eisaki, S.\ Uchida, and J.\ C.\ Davis,
Nature {\bf 411}, 920 (2001).

\bibitem{suc_01} S. Uchida,
Physica C {\bf 357}, 25 (2001).

\bibitem{spa_03} S. H. Pan, E. W. Hudson, K. M. Lang, H. Eisaki, S. Uchida,
and J. C. Davis,
Nature {\bf 403}, 746 (2003).

\bibitem{scheer} E.\ Scheer, N.\ Agra\"{\i}t, J.\ C.\ Cuevas, A.\ Levy Yeyati,
B.\ Ludoph, A.\ Mart\'{\i}n-Rodero, G.\ Rubio Bollinger, J.\ M.\ van Ruitenbeek,
and C.\ Urbina,
Nature {\bf 394}, 154 (1998).

\bibitem{cuevas} J.\ C.\ Cuevas, A.\ Levy Yeyati, A.\ Mart\'{\i}n-Rodero,
G.\ R.\ Bollinger, C.\ Untiedt, and N.\ Agra\"{\i}t,
\prl {\bf 81}, 2990 (1998).

\bibitem{post} N.\ van der Post, E.\ T.\ Peters, I.\ K.\ Yanson, and
J.\ M.\ van Ruitenbeek,
\prl {\bf 73}, 2611 (1994).

\bibitem{post_2} E.\ N.\ Bratus', V.\ S.\ Shumeiko, and
G.\ Wendin,
\prl {\bf 74}, 2110 (1995).

\bibitem{tinkhambook} M. Tinkham, {\it Introduction to Superconductivity},
(McGraw-Hill, Inc., New York, 1996).

\bibitem{ofi_07} \O. Fischer, M. Kugler, I. Maggio-Aprile, C. Berthold, and
C. Renner,
Rev.\ Mod.\ Phys.\ {\bf 79}, 353 (2007).

\bibitem{sosolik} C.\ E.\ Sosolik, J.\ A.\ Stroscio, M.\ D.\ Stiles,
E.\ W.\ Hudson, S.\ R.\ Blankenship, A.\ P.\ Fein, and R.\ J.\ Celotta,
\prb {\bf 68}, 140503(R)(2003).

\bibitem{nbe_08} N. Bergeal, Y. Noat, T. Cren, Th.\ Proslier, V. Dubost,
F. Debontridder, A. Zimmers, D. Roditchev, W. Sacks, and J. Marcus,
\prb {\bf 78}, 140507(R) (2008).

\bibitem{muller} J.\ Muller,
Rep. Prog. Phys. {\bf 43}, 641 (1980).

\bibitem{hardy} G.\ F.\ Hardy and J.\ K.\ Hulm,
Phys.\ Rev.\ {\bf 5}, 1004 (1954).

\bibitem{vonsovsky} S.\ V.\ Vonsovsky, Yu.\ A.\ Izyumov, and E.\ Z.\ Kurmaev,
{\em Superconductivity of Transition Metals}, (Springer-Verlag, Berlin, 1982).

\bibitem{clogston} A.\ M.\ Clogston, A.\ C.\ Gossard, V.\ Jaccarino, and
Y.\ Yafet,
\prl {\bf 9}, 262 (1962).

\bibitem{morin} F.\ J.\ Morin and J.\ P.\ Maita,
Phys.\ Rev.\ {\bf 129}, 1115 (1963).

\bibitem{williams} H.\ J.\ Williams and R.\ C.\ Sherwood,
Bull.\ Am.\ Phys.\ Soc. {\bf 5}, 430 (1960).

\bibitem{paduani} C.\ Paduania and C.\ A.\ Kuhnen,
Eur.\ Phys.\ J.\ B {\bf 66}, 353 (2008).

\bibitem{testardi} L.\ R.\ Testardi,
Rev.\ Mod.\ Phys.\ {\bf 47}, 637 (1975).

\bibitem{yet1} M. Yethiraj, D.\ K.\ Christen, D.\ McK. Paul, P. Miranovic,
and J.\ R.\ Thompson,
\prl {\bf 82}, 5112 (1999).

\bibitem{kogan} V.\ G.\ Kogan, P. Miranovic, Lj. Dobrosavljevic-Grujic,
W.\ E.\ Pickett, and D.\ K.\ Christen,
\prl {\bf 79}, 741 (1997).

\bibitem{sonier} J.\ E.\ Sonier, F.\ D.\ Callaghan, R.\ I.\ Miller, E. Boaknin,
L.\ Taillefer, R.\ F.\ Kiefl, J.\ H.\ Brewer, K.\ F.\ Poon, and J.\ D.\ Brewer,
\prl {\bf 93}, 017002 (2004).

\bibitem{yethiraj} M. Yethiraj, D.\ K.\ Christen, A.\ A.\ Gapud, D.\ McK.Paul,
S.\ J.\ Crowe, C.\ D.\ Dewhurst, R.\ Cubitt, L.\ Porcar, and A.\ Gurevich,
\prb {\bf 72}, 060504(R) (2005).

\bibitem{zajac} G.\ Zajac, J.\ Zak, and S.\ D.\ Bader,
\prb {\bf 27}, 6649 (1983).

\bibitem{aono} M.\ Aono, F.\ J.\ Himpsel, and D.\ E.\ Eastman,
Solid State Commun.\ {\bf 39}, 225 (1981).

\bibitem{nefyodov} Yu.\ A.\ Nefyodov, A.\ M.\ Shuvaev, and M.\ R.\ Trunin,
Europhys.\ Lett.\ {\bf 72}, 638 (2005).

\bibitem{levinstein} H.\ J.\ Levinstein and J.\ E.\ Kunzler,
Phys.\ Lett.\ {\bf 20}, 581 (1966).

\bibitem{hauser} J.\ J.\ Hauser, D.\ D.\ Bacon, and W.\ H.\ Haemmerle,
Phys.\ Rev.\ {\bf 151}, 296 (1966).

\bibitem{schumaun} J. Schumaun and D. Elefant,
Phys.\ Status Solidi B {\bf 95}, 91 (1979).

\bibitem{moore} D.\ F.\ Moore, R.\ B.\ Zubeck, J.\ M.\ Rowell, and M.\ R.\ Beasley,
\prb {\bf 20}, 2721 (1979).

\bibitem{bangert} W. Bangert, J. Geerk, and P. Schweiss,
\prb {\bf 31}, 6066 (1985).

\bibitem{tanner} D.\ B.\ Tanner and A.\ J.\ Sievers,
\prb {\bf 8}, 1978 (1973).

\bibitem{hackl} R. Hackl, R. Kaiser, and W. Gl\"aser,
Physica C {\bf 162-164}, 431 (1989).

\bibitem{reinert} E. Reinert, G. Nicolay, S. H\"ufner, U. Probst, and
E.\ Bucher,
J.\ Electron Spectrosc.\ Relat.\ Phenom.\ {\bf 114-116}, 615 (2001).

\bibitem{suhlbcs} H.\ Suhl, B.\ T.\ Matthias, and L.\ R.\ Walker,
\prl {\bf 3}, 552 (1959).

\bibitem{moskalenko} V.\ Moskalenko,
Fiz.\ Met.\ Metalloved.\ {\bf 8}, 503 (1959).

\bibitem{bcs} J.\ Bardeen, L.\ N.\ Cooper, and J.\ R.\ Schrieffer,
Phys.\ Rev.\ {\bf 108}, 1175 (1957).

\bibitem{chang} C.\ C.\ Chang, Surf.\ Sci.\ {\bf 25}, 53 (1971).

%Auger
\bibitem{auger} L.\ E.\ Davis, N.\ C.\ MacDonald, P.\ W.\ Palmberg,
G.\ E.\ Riach, and R.\ E.\ Weber,
{\em Handbook of Auger Spectroscopy}, (Physical Electronics Industries).

%SiC surface
\bibitem{derycke} V. Derycke, P. Soukiassian, A. Mayne, G.\ Dujardin, and
J.\ Gautier,
\prl {\bf 81}, 5868 (1998).

\bibitem{yu} V.\ Yu.\ Aristov, P.\ Soukiassian, A.\ Catellani, R.\ DiFelice,
and G.\ Galli,
\prb {\bf 69}, 245326 (2004).

\bibitem{tejeda} A. Tejeda, E. Wimmer, P. Soukiassian, D.\ Dunham, E.\ Rotenberg,
J.\ D.\ Denlinger, and E.\ G.\ Michel,
\prb {\bf 75}, 195315 (2007).

\bibitem{bode} M. Bode, R. Pascal, and R. Wiesendanger,
Surf.\ Sci.\ {\bf 344}, 185 (1995).

\bibitem{clogston1} A.\ M.\ Clogston and V.\ Jaccarino,
Phys.\ Rev.\ {\bf 121}, 1357 (1961).

\bibitem{wegner} M. Wegner,
Rev.\ Mod.\ Phys. {\bf 36}, 175 (1964).

\bibitem{labbe} J.\ Labb\'e and J.\ Friedel,
J.\ Phys.\ Radium {\bf 27}, 153 (1966).

\bibitem{mattheiss} L.\ F.\ Mattheiss,
Phys.\ Rev.\ {\bf 138}, A112 (1965).

\bibitem{mattheiss_2} L.\ F.\ Mattheiss,
\prb {\bf 12}, 2161 (1975).

\bibitem{klein} B.\ M.\ Klein, L.\ L.\ Boyer, D.\ A.\ Papaconstantopoulos,
and L.\ F.\ Mattheiss,
\prb {\bf 18}, 6411 (1978).

\bibitem{pui} P.\ K.\ Lam and M.\ L.\ Cohen,
\prb {\bf 23}, 6371 (1981).

\bibitem{mattheiss_3} L.\ F.\ Mattheiss and W.\ Weber,
\prb {\bf 25}, 2248 (1982).

\bibitem{bisi} O. Bisi and L.\ W.\ Chiao,
\prb {\bf 25}, 4943 (1982).

\bibitem{gpa_06} G. Panaccione, J. Fujii, I. Vobornik, G. Trimarchi, N. Binggeli,
A. Goldoni, R. Larciprete, and G. Rossi,
\prb {\bf 73}, 035431 (2006).

\bibitem{gtr_06} G. Trimarchi and N. Binggeli,
Phys.\ Stat.\ Sol.\ B {\bf 243}, 2105 (2006).

%scr
\bibitem{poole} C.\ P.\ Poole, H.\ A.\ Farach, and R.\ J.\ Creswick,
{\em Superconductivity}, (Academic Press, New York, 1995).

\bibitem{eliashberg} G.\ M.\ Eliashberg,
Zh.\ Eksp.\ Teor.\ Fiz.\ {\bf 38}, 966 (1960).

\bibitem{dynes} R.\ C.\ Dynes, V. Narayanamurti, and J.\ P.\ Garno,
\prl {\bf 41}, 21, 1509 (1978).

\bibitem{alexandrov} A.\ S.\ Alexandrov, {\em Theory of \scon{} - from weak to strong coupling}, (Institute of Physics Publishing, Bristol, 2003)

\bibitem{klein_inst} J.\ Klein, A.\ L\'eger, M.\ Belin, D.\ D\'efourneau,
and M.\ J.\ L.\ Sangster,
\prb {\bf 7}, 2336 (1973).

\bibitem{kroeger} J.\ Kr\"{o}ger, L.\ Limot, H.\ Jensen, R.\ Berndt, S.\ Crampin,
and E.\ Pehlke,
Prog.\ Surf.\ Sci.\ {\bf 80}, 26 (2005).

%contact spectroscopy
\bibitem{lli_05} L. Limot, J. Kr\"{o}ger, R. Berndt, A. Garcia-Lekue, and
W. A. Hofer,
\prl {\bf 94}, 126102 (2005).

\bibitem{nne07a} N. N\'{e}el, J. Kr\"{o}ger, L. Limot, K. Palotas, W. A. Hofer,
and R. Berndt,
\prl {\bf 98}, 016801 (2007).

\bibitem{jkr_07} J. Kr\"{o}ger, H. Jensen, and R. Berndt,
New J.\ Phys.\ {\bf 9}, 153 (2007).

\bibitem{jkr_08} J. Kr\"{o}ger, N. N\'{e}el, and L. Limot,
J.\ Phys.: Condens.\ Matter {\bf 20}, 223001 (2008).

\bibitem{simmons} J.\ G.\ Simmons,
J.\ Appl.\ Phys.\ {\bf 34}, 1793 (1963).

\bibitem{lang} N.\ D.\ Lang,
\prb {\bf 37}, 10395 (1988).

\bibitem{olesen} L.\ Olesen, M.\ Brandbyge, M.\ R.\ S\o rensen,
K.\ W.\ Jacobsen, E.\ L\ae gsgaard, I.\ Stensgaard, and F.\ Besenbacher,
\prl {\bf 76}, 1485 (1996).

\bibitem{michaelson} H.\ B.\ Michaelson,
J.\ Appl.\ Phys.\ {\bf 48}, 4729 (1977).

\bibitem{heiniger} F. Heiniger and L. Walld\'{e}n,
Phys.\ Status Solidi {\bf 5}, 82 (1971).

%andreev
\bibitem{nne_08} N. N\'{e}el, J. Kr\"{o}ger, L. Limot, and R. Berndt,
Nano Lett.\ {\bf 8}, 1291 (2008).

\bibitem{nne_09} N. N\'{e}el, J. Kr\"{o}ger, R. Berndt, and E. Pehlke,
\prb {\bf 78}, 233402 (2008).

\bibitem{nne07b} N. N\'{e}el, J. Kr\"{o}ger, L. Limot, T. Frederiksen,
M. Brandbyge, and R. Berndt,
\prl {\bf 98}, 065502 (2007).

\bibitem{gsc_08} G. Schulze, K. J. Franke, A. Gagliardi, G. Romano, C. S. Lin,
A. L. Rosa, T. A. Niehaus, Th.\ Frauenheim, A. Di Carlo, A. Pecchia, and
J. I. Pascual,
\prl {\bf 100}, 136801 (2008).

\bibitem{benistant} P.\ A.\ M.\ Benistant, H.\ van Kempen, and P.\ Wyder,
\prl {\bf 51}, 817 (1983).

\bibitem{soulen} R.\ J.\ Soulen Jr., J.\ M.\ Byers, M.\ S.\ Osofysky,
B.\ Nadgorny, T.\ Ambrose, S.\ F.\ Cheng, P.\ R.\ Broussard, C.\ T.\ Tanaka,
J.\ Nowak, J.\ S.\ Moodera, A.\ Barry, and J.\ M.\ D.\ Coey,
Science {\bf 282}, 85 (1998).

\bibitem{strijkers} G.\ J.\ Strijkers, Y.\ Ji, F.\ Y.\ Yang, C.\ L.\ Chien, and J.\ M.\ Byers,
\prb {\bf 63}, 104510 (2001).

\bibitem{andreev} A.\ F.\ Andreev,
Zh.\ Eksp.\ Teor.\ Fiz.\ {\bf 46}, 1823 (1964).

\bibitem{blonder} G.\ E.\ Blonder, M.\ Tinkham, and T.\ M.\ Klapwijk,
\prb {\bf 25}, 4515 (1982).

\bibitem{sharvin} Yu.\ V.\ Sharvin,
Zh.\ Eksp.\ Teor.\ Fiz.\ {\bf 48}, 984 (1965).

\end{thebibliography}
\end{document}